\documentclass[final,3p,twocolumn]{elsarticle}
\usepackage{graphicx,amsmath,amssymb}

\journal{Physica B}

\begin{document}
\begin{frontmatter}
\title{The Vollhardt crossing point at high magnetic field}
\author{A.E. Petrova}
\author{S.M. Stishov\corref{cor1}\fnref{fn1}}
\address{Institute for High Pressure Physics of RAS, Troitsk, Russia}
\ead{sergei@hppi.troitsk.ru}

\begin{abstract}
Direct measurements of the Vollhardt crossing point coordinates were performed making use data of dilatometric and ultrasound experiments on MnSi. As is shown the crossing points are not invariant in the extended range of magnetic field and should be viewed as a side effects of flattening and spreading out the fluctuation maxima or minima by magnetic field~\cite{Sti16}. Correspondingly the crossing point can not be identified with a characteristic feature controlling the phase transition. So further studies are needed to understand intriguing features of the phase diagram of helical itinerant magnet MnSi.      

\end{abstract}


\end{frontmatter}

\begin{figure}[htb]
\includegraphics[width=80mm]{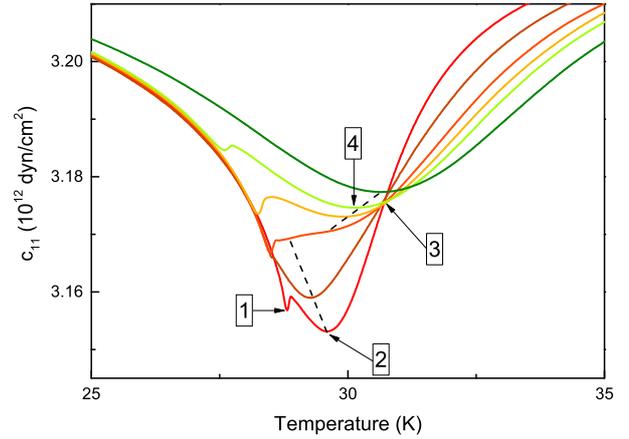}
\caption{\label{fig1} Schematic view of the crossing phenomena in elastic modulus of MnSi as functions of temperature and magnetic field. Numerical labels correspond to features, shown in Fig.4.  Drawn after the data in Ref.~\cite{Pet15}.}
\end{figure}

Twenty years ago D. Vollhardt observed that in some strongly correlated systems the heat capacity curves, measured at different magnetic fields, cross at a single invariant point~\cite{Vollhardt}. Later the crossing effects were observed also in a number of different experimentally obtained quantities~\cite{Greger}. This phenomenon can be seen in the heat capacity, thermal expansion, elastic moduli and sound adsorption in MnSi~\cite{Sti07, Sti08, Pet09, Sti16}. The schematic view of behavior of elastic modulus $c_{11}$ at the phase transition in MnSi as a function of magnetic field is shown in Fig.~\ref{fig1}.

\begin{figure}[htb]
\includegraphics[width=80mm]{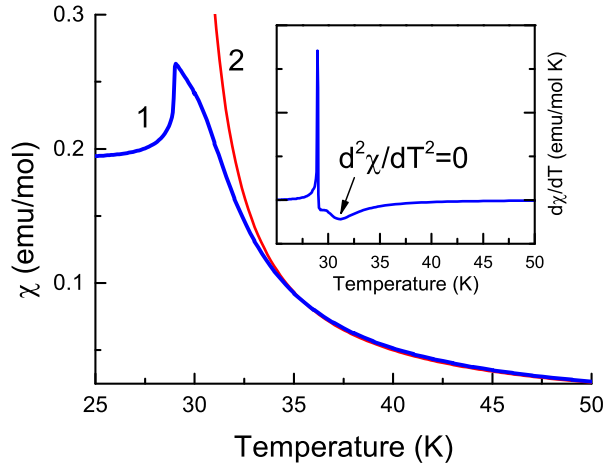}
\caption{\label{fig2} Magnetic susceptibility $\chi=M/H$ of MnSi as a function of temperature (1). Extrapolation of high temperature data according to the Curie-Weiss law (2). Inset: Temperature derivative of magnetic susceptibility $d\chi/dT$ as a function of temperature. Zero value of $d^2\chi/dT^2$ is clearly seen at $\sim 31$ K. This may indicate a change in the correlation length behavior~\cite{Janoschek} (drawn after data of Ref.~\cite{Sti07}).}
 \end{figure}
\begin{figure}[htb]
	\includegraphics[width=80mm]{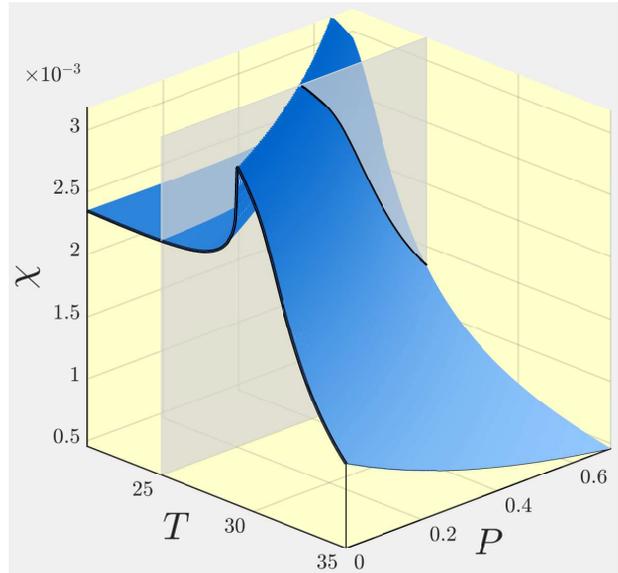}
	\caption{\label{fig3} 3D view of magnetic susceptibility of MnSi as functions of temperature and pressure. Drawn after data in Ref.~\cite{Pet09a}. As is seen an inflection point can be also observed in  $\chi(P)$ dependence and hence in $\chi(P,T)$ as arbitrary function of temperature and pressure.}
\end{figure}

\begin{figure}[!hb]
	\includegraphics[width=80mm]{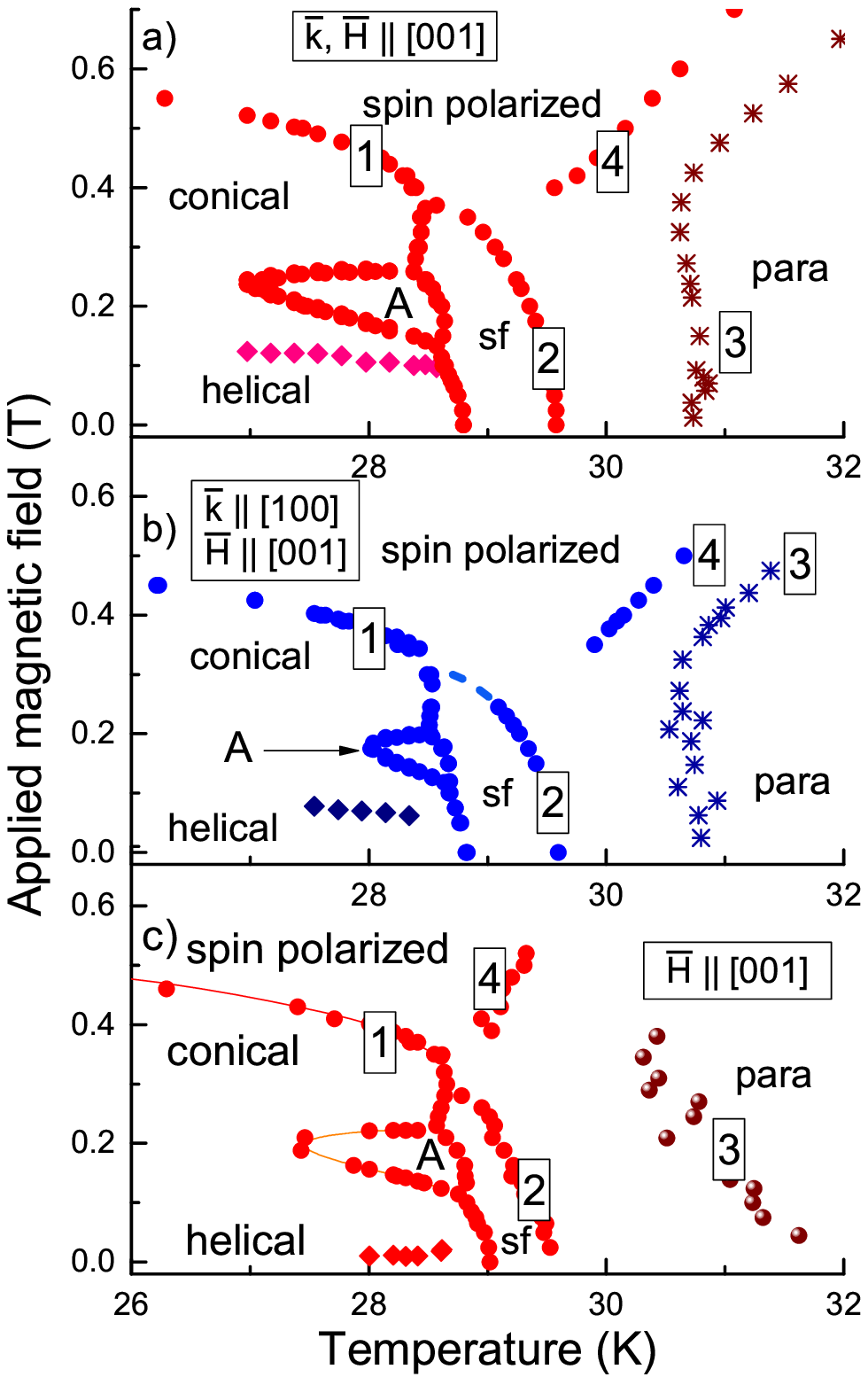}
	\caption{\label{fig4} Magnetic phase diagram of MnSi after ultrasound and dilatometric measurements~\cite{Pet15,Pet16}. 1 - phase transition line, 2 - maximum of the shoulder, 3 - crossing point line, 4 - pseudo transition line between spin polarized and paramagnetic states. $A$ - skyrmion phase. As is seen the crossing points are not invariant under magnetic fields.}
\end{figure}

Incidently authors of~\cite{Sti07} did not recognize any kind of significance of the observed crossing at that time. Later the heat capacity crossing point in MnSi was confirmed in Refs.~\cite{Bauer,Janoschek}. Moreover, as was concluded in Ref.~\cite{Janoschek}, an existence of the invariant crossing point in MnSi followed from the Brazovsky model of the fluctuation induced phase transition in MnSi~\cite{Brazovskii} and signifies a drastic change in the behavior of correlation length.

However, the existence of the invariant crossing points was disputed in Ref.~\cite{Sti16}. One should know that an existence the mentioned invariant crossing points in magnetic fields puts certain restrictions on a behavior of the magnetic susceptibility at the crossing temperature. Indeed, making use the Maxwell relations one can get the following relationship on a condition of linear approximation  $\chi=dM/dH=M/H$.
\begin{equation}\label{eq1}
\left( \dfrac{\partial C_{H}}{\partial H}\right)_{T^*}=T\left(\dfrac{\partial ^{2}M}{\partial T^{2}} \right) _{T^{*}}\approx TH\left( \dfrac{\partial^{2} \chi}{\partial T^{2}}\right) _{T^{*}}=0 
\end{equation}
\begin{equation}\label{eq2}
\left( \dfrac{\partial \beta}{\partial H}\right)_{T^*}=\dfrac{1}{V} \left( \dfrac{\partial ^{2}M}{\partial ^{2}p} \right)_{T^*}\approx\dfrac{H}{V}\left( \dfrac{\partial ^{2}\chi}{\partial ^{2}p} \right)_{T^*}=0
\end{equation}
\begin{equation}\label{eq3}
\left( \dfrac{\partial \alpha}{\partial H}\right) _{T^{*}}=-\dfrac{1}{V}\left(\dfrac{\partial ^{2}M}{\partial p \partial T} \right) _{T^{*}}\approx -\dfrac{H}{V}\left( \dfrac{\partial^{2}\chi}{\partial p \partial T}\right) _{T^{*}}=0
\end{equation}

Here $C_{H}$ - heat capacity, $M$ - magnetization, $\chi=M/H$ - magnetic susceptibility, $\beta=-{1}/{V}( {\partial V}/{\partial p})_{T}$ - compressibility, $\alpha=(1/V) (\partial  V /  \partial T)_{P}$ - thermal expansion coefficient.

The eq.~\ref{eq1} can be validated in Fig.~\ref{fig2}, which demonstrates an inflection point on the $\chi (T)$ curve in MnSi. There are no direct proofs for an existence of the inflection point on the $\chi (P)$ curve (eq.~\ref{eq2}), but a simple modeling the situation confirms that should be a case (see Fig.~\ref{fig3}).  At the same time Fig.~\ref{fig3} can be served as a verification of eq.~\ref{eq3} as well.
However, the zero value of $d^2\chi/dT^2$ seen in Fig.~\ref{fig2} does not imply an existence of the invariant crossing point. Indeed, the equations (1-3) are always will be valid for any pair of corresponding crossing curves. Hence it is important to see whether the relationship \eqref{eq1}--\eqref{eq3} can be justified in an extended range of magnetic fields, where  the approach $\chi =M/H$ is hold, with a number of crossing curves.

Direct measurements of the crossing point coordinates are shown in Fig.~\ref{fig4}. The corresponding data inserted in the magnetic phase diagrams of MnSi, obtained in the dilatometric and ultrasound experiments~\cite{Pet15, Pet16}. As can be seen from the diagrams the crossing points are not invariant in the extended range of magnetic field and should be viewed as a side effects of flattening and spreading out the fluctuation maxima by magnetic field~\cite{Sti16}. This conclusion makes it clear that the puzzling features of the phase diagram of the itinerant magnet MnSi are not well understood and further studies are needed.

\section{Conclusion}
Direct measurements of the Vollhardt crossing point coordinates were performed making use data of dilatometric and ultrasound experiments on MnSi. As is shown the crossing points are not invariant in the extended range of magnetic field and arise as a result of flattening and spreading out the fluctuation maxima or minima by magnetic field. Correspondingly the crossing points can not be identified with a characteristic feature controlling a phase transition~\cite{Sti16}. So further studies are needed to understand intriguing features of the phase diagram of helical itinerant magnet MnSi.      

\section{Acknowledgements}
This work was supported by the Russian Foundation for Basic Research (grant No. 18-02-00183), the Russian Science foundation (grant 17-12-01050) and Program of the Physics Department of RAS on Strongly Correlated Electron Systems and Program of the Presidium of RAS on Strongly Compressed Matter.


\begin{thebibliography}{99}
\bibitem{Vollhardt} D. Vollhardt, Phys. Rev. Lett. {\bf 78} (1997) 1307.
\bibitem{Greger} M. Greger, M. Kollar, D. Vollhardt, Phys.Rev. B{\bf 87} (2013) 195140.
\bibitem{Sti07}	S.M. Stishov, A.E. Petrova, S. Khasanov, G.Kh. Panova, A.A. Shikov, J.C. Lashley, D. Wu, T.A. Lograsso, Phys.Rev. B{\bf 76} (2007) 052405.
\bibitem{Sti08}	S.M. Stishov, A.E. Petrova, S. Khasanov, G.Kh. Panova, A.A. Shikov, J.C. Lashley, D. Wu, T.A. Lograsso, J. Phys. Condens. Matter, {\bf 20} (2008) 235222.
\bibitem{Pet09}	A.E. Petrova, S.M. Stishov, J. Phys.: Condens. Matter, {\bf 21}  (2009) 196001.
\bibitem{Sti16} S.M. Stishov, A.E. Petrova, Phys. Rev. B {\bf 94} (2016) 140406(R).
\bibitem{Bauer} A. Bauer, A. Neubauer, C. Franz, W. M\"{u}nzer, M. Garst, C. Pfleiderer, Phys.Rev. B{\bf 82} (2010) 064404.
\bibitem{Janoschek} M. Janoschek, M. Garst, A. Bauer, P. Krautscheid, R. Georgii, P. B\"{o}ni, C. Pfleiderer, Phys.Rev. B {\bf 87} (2013) 134407.
\bibitem{Pet15} A.E. Petrova, S.M. Stishov, Phys.Rev., B {\bf 91} (2015) 214402.
\bibitem{Pet16} A.E. Petrova, S.M. Stishov, Phys.Rev., B {\bf 94} (2016) 020410(R).
\bibitem{Brazovskii} S. A. Brazovskii, Zh. Eksp. Teor. Fiz. {\bf 68}, 175 (1975).
\bibitem{Pet09a} A. E. Petrova, V. N. Krasnorussky, T. A. Lograsso, S. M. Stishov, Phys.Rev. B {\bf 79}, 100401 (R) (2009)



\end{thebibliography}
\end{document}